\documentclass[twocolumn,showpacs,preprintnumbers,amsmath,amssymb,floatfix,
prl]{revtex4}

\usepackage{graphicx}
\usepackage{dcolumn}
\usepackage{bm}
\usepackage{epsfig}
\usepackage{amsmath}
\usepackage{ulem}
\usepackage{color}

\newcommand{\be}{\begin{equation}}
\newcommand{\ee}{\end{equation}}
\newcommand{\beq}{\begin{eqnarray}}
\newcommand{\enq}{\end{eqnarray}}

\begin{document}


\title{First and second sound in cylindrically trapped gases } 

\date{\today}

\author{G. Bertaina$^{1}$}
\author{L. Pitaevskii$^{1,2}$ }
\author{S. Stringari$^{1}$}
\affiliation{$^{1}$INO-CNR BEC Center
 and Dipartimento~di~Fisica, Universit\`a di Trento, I-38123 Povo, Trento, Italy\\
$^{2}$Kapitza Institute for Physical Problems, Kosygina 2, 119334 Moscow, Russia}

\begin{abstract}

We investigate the propagation of density and temperature waves in a cylindrically trapped gas with radial harmonic confinement. Starting from  two-fluid hydrodynamic theory  we derive effective 1D equations for  the chemical potential and the temperature which explicitly account for the effects of viscosity and thermal conductivity. Differently from quantum fluids confined by rigid walls, the harmonic confinement allows for the propagation of both first and second sound in the long wave length limit. We provide quantitative predictions for the two sound velocities of a superfluid Fermi gas at unitarity. For shorter wave-lengths we discover a new surprising class of excitations continuously spread over a finite interval of frequencies. This results in a non-dissipative damping in the response function which is analytically calculated in the limiting case of a classical ideal gas.

\end{abstract}


\maketitle

Superfluids are known to exhibit, in addition to 
usual sound, an additional mode, called second sound,  where the
superfluid and normal components  move with
opposite phase \cite{landau}. In a weakly compressible fluid, like
superfluid helium, second sound reduces to 
a temperature wave, leaving the total density practically
unaffected. In helium its velocity was systematically measured as a function
of temperature, providing a high precision determination
of the superfluid density. First measurements of second
sound were recently reported also in dilute Bose-Einstein
condensed gases confined in elongated traps [2].
In this Letter we show that the propagation of 
sound in a nonuniform trapped gas exhibits new interesting
features both in the superfluid and in the normal phase. We will focus on highly elongated
configurations where the radial harmonic trapping
provides the relevant non-uniformity. These configurations
are well suited for the experimental excitation
and detection of sound waves \cite{ketterle,thomas,ueda}. We will consider the usual
hydrodynamic regime $l \ll \lambda$ where $l$ is the mean free
path and $\lambda $ is the wavelength of the sound wave. We
will also consider the condition of strong radial confinement
$R_{\perp} \ll \lambda $ where $R_{\perp}$ is the radial size of the sample.

The above conditions are compatible
with two different regimes, depending on the ratio
between the viscous penetration depth $\delta =\sqrt{\eta/\rho_{n}\omega}$
 and the radius $R_{\perp}$ where $\eta$ is the shear viscosity coefficient,
$\rho_{n}$ is the normal density and $\omega$ is the frequency of the
sound wave. Let us first consider the case of a uniform
fluid confined by the hard walls of a tube. If $\delta \ll R_{\perp}$
viscosity plays an unimportant role and 
sound propagates similarly to the case of bulk matter. If instead $\delta \gg R_{\perp}$, the viscosity
imposes the uniformity of the normal velocity field as
a function of the radial coordinate.  Since friction  further requires that
the normal velocity be zero on the walls, the normal part of the fluid cannot move 
at all along the tube and only the superfluid
can. This mode is known as 4-th
sound \cite{atkins}. In the case of a harmonically trapped gas 
the two  regimes exhibit new features whose investigation is the
object of the present paper. 

Our analysis is based on the two-fluid hydrodynamic equations (HD) \cite{khalatnikov}
\begin{align}
&\partial_t \rho
+ \bm{\nabla}\cdot\mathbf{j} = 0, \,
m\partial_t \mathbf{v}_s
 =-\bm{\nabla}\left(\mu + U\right), \label{cont}\\
&\partial_tj_i
+ \partial_i P + \rho\partial_i U/m 
=\partial_k (\eta \Gamma_{ik}),  \label{dtj} \\
&\partial_t s + \bm{\nabla}\cdot(s\mathbf{v}_n) = 
\mathbf{\nabla}\cdot\left (\kappa\mathbf{\nabla}T/T\right), %
\label{st}
\end{align}
which reduce, above $T_{C}$,
to the classical equations of hydrodynamics.
In the above equations  $\mathbf{j} = \rho_s\mathbf{v}_s+\rho_n\mathbf{v}_n$ is the current
density, $\rho_s$ and $\rho_n$ are the superfluid and normal mass densities
for a fluid with total mass density $\rho = \rho_s+\rho_n$, ${\mathbf v}_s$ and ${\mathbf v}_n$ are the corresponding velocity fields, 
$\Gamma_{ik}= (\partial_k v_{ni}+\partial_i v_{nk}-2\delta_{ik}\partial_j v_{nj}/3)$, $s$ is the entropy density, while $P$ and $\mu$ are, respectively, the local pressure and chemical potential.
 $U$ is the external trapping potential which will be assumed of axial harmonic form: $U = m(\omega^2_\perp r^2_\perp+\omega^2_zz^2)/2$ with $r^2_\perp = x^2+y^2$ and $\omega_z \ll \omega_\perp$. Since
we are interested only in  the linear solutions,  in the above equations we have omitted terms quadratic in the velocity.  
Finally $\eta$ and $\kappa$ are, respectively, the shear viscosity and the thermal conductivity (we have not considered  bulk viscosities terms which give smaller contributions).   
In a uniform fluid the effect of viscosity and thermal conductivity is irrelevant in the long wave length limit, providing only higher order corrections to the dispersion law. 
In the presence of confinement their effect is instead crucial. This opens new perspectives to explore the behavior of viscosity and thermal conductivity in interacting trapped gases.
The two-fluid hydrodynamic equations, in the dissipationless
regime ($\eta= \kappa=0$), have been already the object of numerical
investigations \cite{Taylor,Taylor2,levin} in the presence of harmonic confinement.  
 These works, 
so far limited to isotropic trapping, have provided predictions
for the discretized frequencies  in the superfluid
phase with special emphasis to the hybridization effects
between the first and second sound solutions. 

In this Letter we are interested in the  solutions of the hydrodynamic equations satisfying the condition $\omega\ll \omega_\perp$. As a consequence of the tight radial confinement, Eq. (\ref{dtj}) for the current  implies the important condition $\nabla_\perp P+\rho \nabla_\perp U=0$ of mechanical equilibrium along the radial direction. Violation of this condition would in fact result in frequencies of the order of $\omega_\perp$.
The tight radial confinement also implies that the  radial component of the velocity field must be much smaller than the longitudinal one. 

\textit{Low frequency regime.}  In the
presence of harmonic trapping the condition  $\delta \gg R_{\perp}$ of large viscous penetration depth 
is equivalent to requiring the low frequency condition $\omega \ll \omega_{\perp}^2\tau $ where
$\tau $  is a typical
collisional time. 

Excluding ${\mathbf v}_s$ from Eqs. (\ref{dtj}) and (\ref{cont}) and using the thermodynamic identity $dP=sdT +\rho d\mu/m$
one can write the equation for the  relevant $z$ component of the velocity field of the normal component in the form
\begin{equation}
m\rho_n \partial_t {v}_n^z
+\rho_n
\partial_z\left(\delta \mu \right) +{ms}\partial_z (\delta T)
= m \bm{\nabla}\cdot 
(\eta\bm{\nabla}v_n^z)\; ,
\label{vn}
\end{equation}
where we have ignored terms containing the  small    radial components of the velocity field. 

The presence of viscosity  in Eq. (\ref{vn}) results in the independence of
${v}_n^z$  on the radial coordinate $r_{\perp}$.  In fact violation of such a behavior would be incompatible with the low frequency condition $\omega \ll \omega_{\perp}^2\tau $. It is worth noticing that, differently from the case of the tube with hard walls discussed in the introduction, for harmonic trapping the uniformity of the velocity field does not stop the motion of the normal component.  Analogously, the presence of thermal conductivity in Eq. (\ref{st}) for the entropy results in the independence of the temperature fluctuations on $r_{\perp}$ in the same low frequency limit \cite{note}. This in turns implies  that also the fluctuations of the chemical potential will be independent of the radial coordinate. This follows from the radial mechanical equilibrium condition   
and the use of the thermodynamic identity  $dP=sdT +\rho d\mu/m$. Thus in the low frequency limit both the fluctuations $\delta T $ and $\delta \mu$ are  independent of the radial coordinates. 

It is now convenient to integrate radially both the equations of continuity for the density and for the entropy (\ref{cont}, \ref{st}) as well as  Eq. (\ref{vn}) for $v^z_n$, profiting of the fact
that the large terms containing the derivatives with respect to $\mathbf{r}_{\perp}$ vanish due to Gauss
theorem, and that  those containing the second derivative with respect to $z$ can be ignored since they give rise to higher order corrections in the dispersion law. Integration of Eq.(\ref{vn}) then gives $-m\langle\rho_n\rangle \partial_t v_n^z=m\langle s \rangle \partial_z \delta T +\langle \rho_n\rangle \partial_z \delta \mu$,
where the symbol $\langle ...\rangle \equiv \int ...dxdy$ stands for the integral on the transverse variable $\mathbf{r}_{\perp}$.
The above equation for $v^z_n$ and the 
analog equation $m\partial_tv^z_s=-\partial_z  \delta \mu$ for the superfluid component can be used to derive the following coupled equations for the averaged density and entropy fluctuations:
\begin{eqnarray}
-\partial^2_t \delta \langle \rho \rangle +  \partial_z\left(\langle s\rangle \partial_z \delta T\right) + 
\partial_z \left(\langle \rho\rangle \partial_z \delta\mu\right)/m =0, 
\label{jt3}\\
-\partial^2_t \delta \langle s\rangle + \partial_z\left(\langle {s}\rangle^2 / \langle {\rho_n}\rangle \partial_z \delta T\right) +
\partial_z \left(\langle{ s }\rangle  \partial_z \delta\mu\right)/m =0.
\label{jt4}
\end{eqnarray}
One can further write 
$\delta\langle \rho \rangle  =  \partial_\mu\langle \rho \rangle \delta\mu +
\partial_T\langle \rho \rangle  \delta T$ and $\delta\langle s\rangle = \partial_\mu\langle s\rangle \delta\mu +\partial_T\langle s\rangle \delta T$ 
The spatial dependence of the  thermodynamic quantities entering the above equations is fixed by the trapping potential $U$ through the local density approximation equilibrium condition $ \mu(\rho,T) + U=\mu_0$, 
 with $\mu_0$ fixed by the normalization condition for $\rho$.
In the presence of axial trapping ($\omega_z\ne 0$), Eqs. (\ref{jt3},\ref{jt4}) allow for the calculation of the low frequency discretized states of the system. In the
following we will assume $\omega_z=0$ (cylindrical geometry) in order to calculate 
the
velocity of the sound waves propagating along the axial direction. The solutions for $\delta \mu$ and
$\delta T$ have the $t$ and  $z$ dependence $\propto e^{i(qz-\omega t)}$ with $\omega=cq$, yielding the most
relevant equation
\begin{align}
c^4 [m\partial_{\mu}\langle \rho \rangle \partial_{T}\langle s \rangle-(\partial_{T}\langle \rho \rangle)^2 ] +
c^2 
[2\partial_{T}\langle \rho \rangle\langle s \rangle-& \nonumber\\
\langle \rho \rangle \partial_{T}\langle s \rangle-m\partial_{\mu}\langle \rho \rangle
\langle s\rangle^2/\langle \rho_n \rangle ] +
\langle \rho_s \rangle \langle s \rangle^2/\langle \rho_n \rangle &=0,
\label{disp}
\end{align}
which generalizes  the well known Landau equation \cite{landau, khalatnikov} for the first and second sound velocities to the case of a cylindrically trapped gas. 

Above $T_{C}$, where the superfluid density vanishes, Eq.(\ref{disp}) admits only one solution with velocity different from zero. 
In this case the value of $c$ can be more conveniently calculated by direct radial integration of the HD Eqs. (\ref{cont}-\ref{st}). In the low frequency limit the equations of continuity for the density and the entropy yield the 1D isoentropic condition $\partial_t(\langle s\rangle/\langle\rho\rangle)=0$ and one immediately finds the result
$c^2 = (\partial \langle P\rangle/ \partial \langle\rho\rangle)_{\langle s\rangle/\langle\rho\rangle}$,
which generalizes the most famous expression for the adiabatic sound velocity of uniform matter as well as the zero temperature result $mc^2 = (\partial_\mu\langle \rho \rangle/\langle \rho \rangle )^{-1}$  derived in \cite{capuzzi06}. At unitarity, as well as for the ideal gas, where $s/\rho$ depends on the combination $\rho T^{-3/2}$, the new adiabaticity condition takes the form $\langle\rho\rangle T^{-5/2}= const$. A non trivial implication concerns the sound velocity in the ideal classical limit where one finds $c^2=(7/5)(T/m)$ rather than the value $(5/3)(T/m)$ holding in uniform gases. Here and in the following we put the Boltzmann constant $k_B=1$.

Before discussing  the solutions of Eq.  (\ref{disp}) in the presence of harmonic trapping, it is useful to calculate the dynamic response function of the system to an external potential  of the form $ \lambda e^{i(qz-\omega t)}$. The response function is determined by the density fluctuations induced by the external potential according to $\delta \langle \rho \rangle = \lambda\chi(q,\omega)e^{i(qz-\omega t)}$. Its  calculation is important because it provides  information on the actual possibility of exciting the two sound waves using a density probe. The inclusion of the external perturbation affects both the equation for the current and the equation for ${\bf v}_s$. 
Calculations yield the following result for the imaginary part of the response function 
\begin{multline}
\rm{Im} \chi(q,\omega)=-\langle \rho \rangle \frac{\pi}{2} \{ W_1 \omega  [\delta(\omega-c_1q)+ \delta(\omega+c_1q)]
 \\
+  W_2 \omega [\delta(\omega-c_2q)+\delta(\omega+c_2q)]\},
\label{Imchi}
\end{multline}
where the weights $W_1$ and $W_2$ obey the relationships  $mc^2_1W_1+mc^2_2W_2=1$ and $W_1 + W_2= (\partial_\mu\langle \rho\rangle /\langle \rho \rangle)_T$  determined, respectively, by the $f$-   and the isothermal compressibility sum rules.   The knowledge of $\rm{Im}\chi$ is relevant for experiments based on the propagation of density pulses and can be  measured 
directly in 
Bragg scattering 
experiments
 \cite{Hu,nikuni}. 

We are now ready to calculate the values for the two sound velocities and the relative weights  in the density response. We will specialize to the case of the unitary Fermi gas \cite{rmp} since in this case the collisional regime, needed to apply  hydrodynamic theory, is more easily achieved due to the large value of the scattering length. At unitarity the effects of the interactions in the
thermodynamic functions can be expressed in a universal way \cite{ho} 
in terms of the  dimensionless parameter  $u\equiv \mu/T$. Thus the pressure can be written as $P(\mu,T)=T^{5/2}H(u)$
where the function $H(u)$  can be determined through microscopic many-body
calculations \cite{Bulgac06,levin2,huDrummond} or, in some ranges
of temperature, directly extracted from experiment \cite{salomon}. All the thermodynamic functions
can be expressed in terms of the function $H$. For example the density takes the form
$\rho(\mu,T)=mT^{3/2}\nu(u)$ with $\nu(u)=H^\prime(u)$, the entropy density is given by $s=5P/2T-u\rho/m$
etc.. Only the superfluid mass density $\rho_s=mT^{3/2}\nu_s(u)$ requires the knowledge of another independent
function for which we use the regularized form introduced in \cite{rhos}.

\begin{figure}[t]
 \includegraphics[width=0.98\columnwidth]{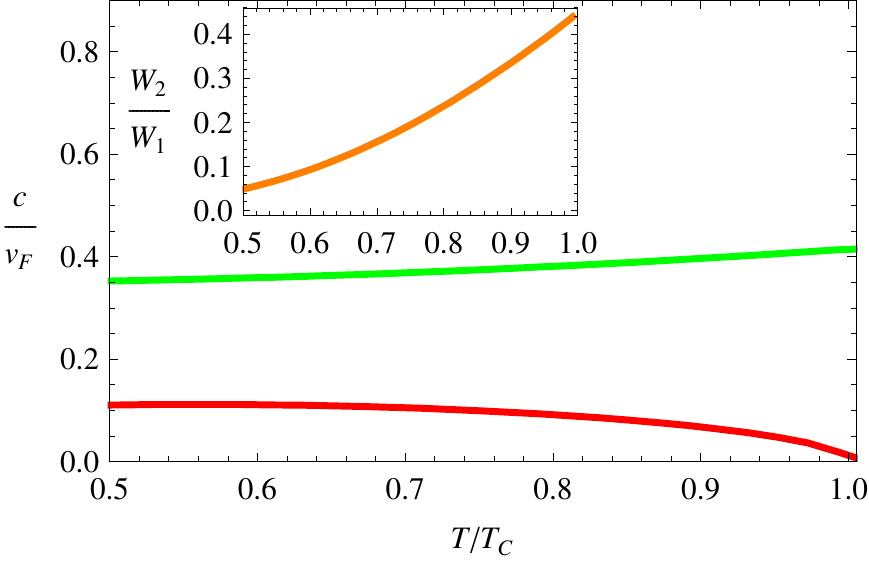}
 \caption{ First (green) and second (red) sound calculated using
Eq.(\ref{disp}), as a function of $T/T_C$, normalized to the Fermi velocity
$v_F=\sqrt{2 T_F/m}$. The inset reports the ratio $W_2/W_1$ between the  weights of the two sounds
modes in ${\rm Im}\chi$ [see Eq. (\ref{Imchi})].}
  \label{fig:velocities}
\end{figure}

\begin{figure}[b]
 \includegraphics[width=0.9\columnwidth]{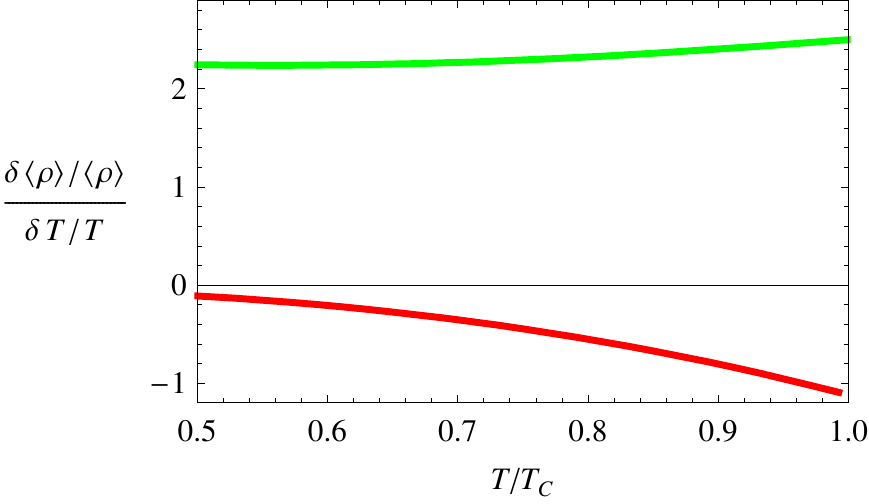}
 \caption{Ratio between the relative temperature ($\delta T/T$) and density ($\delta \langle \rho \rangle /\langle \rho \rangle$) fluctuations in the two sounds as a function of  $T/T_C$ (see Fig. \ref{fig:velocities} for the notation).}
\label{fig:ratio}
\end{figure}

The radially integrated quantities entering the dispersion relation (\ref{disp})  can also be easily
calculated. Since in the presence of radial trapping the chemical potential has the radial dependence
$\mu({\bf r})= \mu_0 -m\omega^2_{\perp}r^2_\perp/2$  the integration over $dxdy$ can be usefully transformed into
integrals over $u$.  For example, we find $\langle \rho \rangle(\mu_0,T)=
(2\pi T^{5/2}/\omega_{\perp}^2)\int_{-\infty}^{\mu_0/T}\nu(u)du$ 
and analogously for the other quantities. In conclusion all the coefficients of the dispersion
relation (\ref{disp}) can be calculated as a function of $T$ and of the dimensionless variable $u_0=\mu_0/T$. 
In practice, in order to determine the function $H(u)$ we have used a fit to the experimental data of 
\cite{salomon} above the critical temperature $T_C$ and assumed that, below $T_C$, the pressure is constant as a function of $T$, following the analysis in the same reference. 
The results for the  sound velocities $c_1$ and $c_2$ and for  the relative weights
$W_2/W_1$ [see (\ref{Imchi})] are reported in Fig. \ref{fig:velocities} for the unitary Fermi gas 
as a function of $T/T_C$ in the relevant interval $0.5 T_C<T<T_C$, where
$T_C\approx 0.19 T_F$, having defined the Fermi temperature for our cylindrical geometry as $T_F =  (\frac{15 \pi }{4} (\frac{\hbar ^2}{2 m})^{3/2} \omega^2_\perp \langle \rho \rangle )^{2/5}$ \cite{fermitemperature}.
In Fig. \ref{fig:ratio} we show the ratio between the relative  density ($\delta \langle \rho \rangle /\langle \rho \rangle$) and temperature ($\delta T/T$) fluctuations in the two modes. 
The ratio provides a physical insight on the nature of the  modes. In the first sound solution (high velocity mode) the relative variations of density are larger than the ones of temperature and have the same sign. 
The opposite happens for second sound.  The value of the ratio $W_2/W_1$ is smaller than the value found for a uniform fluid \cite{Hu} and this is due to the reduction of the superfluid component as one moves from the symmetry axis of the trap. We have also checked that our predictions are not very sensitive to the fitting procedure used to calculate the thermodynamic functions.

\textit{High frequency regime}. We have so far discussed the solutions of the hydrodynamic equations in the 
low frequency limit. In the opposite limit $\omega \gg \omega_\perp^2\tau$ of large frequencies viscosity and thermal conductivity 
can be ignored \cite{isotropic}. In this limit, still compatible with the
condition $\omega \ll \omega_\perp $, the solutions  are also affected by the presence of the radial confinement, but in  a different way. 
In particular the fluctuations of the temperature and of the chemical potential are no longer  independent of the radial coordinate, except in the superfluid region \cite{separate}. 
This limit is still compatible with the usual hydrodynamical condition $\omega\tau \ll 1$ only if $\omega_{\perp}\tau \ll 1$, a condition which is rather severe from the experimental point of view. 

An explicit and instructive  solution  in the dissipationless regime is available 
above $T_{C}$ where the superfluid component is absent. We will consider the  extreme  regime of the ideal gas where all the thermodynamic quantities are known explicitly. In this limit the classical equations of hydrodynamics can be recast in the form of an
equation for the velocity field \cite{GWS}:
\begin{equation}
-m \omega^2\mathbf{v} ={\frac{5}{3}}{T}
{\nabla }[{\nabla }\cdot \mathbf{v}]-{\nabla }[\mathbf{v}\cdot {\nabla }U]  
-{\frac{2}{3}}[{\nabla }\cdot \mathbf{v}]{\nabla }
U  \;.
 \label{GWS}
\end{equation}
Solutions of this equations for 3D harmonic traps were considered in \cite{BC} and \cite{CG}.
The advantage of using 1D cylindrical configurations is due to the availability of plane wave solutions along the $z$ direction, providing new physical insight on the effects of trapping on sound propagation.
It was found in \cite{griffinAd} that an exact solution of this equation in the presence of cylindrical radial trapping ($U = m\omega^2_\perp r^2_\perp/2)$ is given by  $v_z \propto e^{\xi^2/5}e^{i q z},v_{\perp}=0 $,
where $\xi^2=m\omega_{\perp}^2r_{\perp}^2/T$, and is characterized by the adiabatic velocity of sound $mc_a^2=5T/3 $. This solution is quite different from the one holding in the low frequency regime discussed in the first part of the paper where $v_z$ does not depend on the radial coordinate. Also the temperature and  the chemical potential fluctuations  
exhibit a strong radial dependence as a consequence of the radial trapping and the absence of thermal conductivity and viscosity. From the explicit knowledge of the velocity field   and  the orthogonality condition $\int \rho \mathbf{v}^*_{(i)}\cdot\mathbf{v}_{(j)} dxdy= 0$ if $\omega_i\neq\omega_j$ 
for the solutions of the HD equations \cite{Taylor2}, it is possible to calculate the contribution of the adiabatic sound  
to the dynamic response function. Using the expression $\rho \propto e^{-\xi^2/2}$ for the density distribution at equilibrium one obtains, after straightforward algebra, the result
\begin{equation}
\chi_{a}(q,\omega)= \frac{5}{9}\frac{\langle \rho \rangle}{m}\frac{q^2}{q^2c_a^2-\omega^2}\;,
\end{equation}
showing that the adiabatic mode does not exhaust the  $f$-sum rule $\chi (\omega\to \infty)= 
-q^2\langle \rho \rangle/m\omega^2$, differently from the case of a uniform gas, and hence revealing that additional solutions of the hydrodynamic Eq. (\ref{GWS}) should exist. 
This  equation actually admits an additional class of low frequency solutions lying in the continuum and consequently exhibiting damping. On the $ \omega-q$ plane these solutions occupy the region 
$\omega/q <\sqrt{24/25}c_a \equiv c_0$. The velocity field  is given by the expressions
$v_z =A e^{\xi^2/4}\left [\sigma \cos (\sigma\xi^2)-\sin (\sigma\xi^2)/20 \right]e^{i q z}$,
 where $\sigma = \sqrt{c_0^2q^2/\omega^2-1}/4$ is a continuous parameter, and $v_{\perp} =Be^{\xi^2/4}\sin (\sigma\xi^2)e^{i q z}/\xi$, with 
 $B=-iA\sqrt{3/20}\left [(c_a^2q^2-\omega^2)/qc_a\omega_{\perp} \right ]$. 
We present here the result for the imaginary part of the dynamic response function in the  region of the continuum 
\begin{equation}
\mathrm{Im}\chi_{c}(q,\omega)= \frac{4}{3}\frac{\langle \rho \rangle }{mc_0^2}\frac{\omega \sqrt{c_0^2q^2-\omega^2}}{c_a^2q^2-\omega^2}\; .
\label{ImchiC}
\end{equation}
It is easy to show that inclusion of the continuum allows for a complete fulfillment of both the compressibility and the $f$-sum rules.
The function $\mathrm{Im}\chi_{c}$ exhibits a sharp maximum at $\omega/q \approx 0.98c_0$. Thus the new excitation can be considered as the analog of second sound
above $T_{C}$. It is worth emphasizing that the damping associated with the continuum is not a consequence of dissipative effects 
as happens in  uniform fluids above $T_{C}$, but a result of the energy leakage from the gas due to the increase of the velocity field on large distances.
The existence of a continuous spectrum of excitations and the value of the limiting velocity $c_0$ are universal properties holding also at lower temperatures, being fixed by the asymptotic behavior at large $r_\perp$ where the gas is classical and ideal.

We thank our collaborators H. Hu, E. Taylor and A. Griffin for important suggestions that led to our work,
and P. Hohenberg, C. Salomon, H. Stoof and P. van der Straten for useful discussions. We also thank the authors of \cite{Bulgac06, huDrummond, salomon, rhos} for providing us with tables of their results.  Financial support from the EuroQUAM Fermix program and from MIUR PRIN is acknowledged.


\begin{thebibliography}{99}

\bibitem{landau} L.~D.~Landau, J. Phys. (U.S.S.R.) {\bf 5}, 71 (1941).

\bibitem{utrecht}  R.~Meppelink, S.~B.~Koller, and P.~van~der~Straten, Phys. Rev. A {\bf 80}, 043605 (2009). 

\bibitem{ketterle} M.~R.~Andrews et al., Phys. Rev. Lett. \textbf{79} 553 (1997).

\bibitem{thomas} J.~Joseph et al., Phys. Rev. Lett. \textbf{98} 170401 (2007).

\bibitem{ueda} M.~Horikoshi et al.,  Science \textbf{327}, 442 (2010).

\bibitem{atkins} K.~R.~Atkins, Phys. Rev. \textbf{113}, 962 (1958). 

\bibitem{khalatnikov} I.~M.~Khalatnikov, \textit{An Introduction to the Theory of
Superfluidity} (Benjamin, New York, 1965).

\bibitem{Taylor} E.~Taylor and A.~Griffin, Phys. Rev. A \textbf{72}, 053630 (2005).

\bibitem{Taylor2} E.~Taylor et al., Phys. Rev. A \textbf{80}, 053601 (2009).

\bibitem{levin} Y.~He, Q.~Chen, C.~C.~Chien, and K.~Levin, Phys Rev. A {\bf 76}, 051602(R) (2007).

\bibitem{note} We  assume here that the collisional times responsible for  viscosity and thermal conductivity 
are comparable.

\bibitem{capuzzi06} P.~Capuzzi et al.,  Phys.
Rev. A \textbf{73}, 021603(R) (2006). 



\bibitem{Hu} H.~Hu et al., New. J. Phys. \textbf{12}, 043040 (2010). 


\bibitem{nikuni} E.~Arahata and T.~Nikuni, Phys. Rev. A {\bf 80}, 043613 (2009).

\bibitem{rmp} S.~Giorgini, L.~Pitaevskii, and S.~Stringari, Rev. Mod. Phys. {\bf 80}, 1215 (2008).

\bibitem{ho} T.~L.~Ho, Phys. Rev. Lett. {\bf 92}, 090402 (2004).

\bibitem{levin2} Q.~J.~Chen, J.~Stajic, and K.~Levin, Phys. Rev. Lett. \textbf{95},
260405 (2005).

\bibitem{Bulgac06} A.~Bulgac, J.~E.~Drut,  and P.~Magierski,   Phys.
Rev. Lett. \textbf{96}, 090404 (2006).

\bibitem{huDrummond} H.~Hu, X.~Liu, and P.~Drummond, New J. Phys. {\bf 12}, 063038 (2010).

\bibitem{salomon} S.~Nascimbene et al., Nature \textbf{463} 1057 (2010).

\bibitem{rhos} N.~Fukushima et al.,  Phys.~Rev.~A \textbf{75}, 033609 (2007); E.~Taylor et al.,  Phys. Rev. A 77, 033608 (2008). 

\bibitem{fermitemperature} Notice that, in the Local Density Approximation regime where $T_F\gg \hbar\omega_\perp$, the ratio $T_C/T_F$ does not depend on the value of the radial confinement.



\bibitem{isotropic} For isotropic trapping the hydrodynamic equations always reduce to the dissipationless regime if $\omega\tau \ll 1$.

\bibitem{separate} In superfluids  the fluctuations $\delta T$ and $\delta\mu$ are independent of the radial coordinate as a consequence of  the equations for $\mathbf{v}_{s}$ and for the radial mechanical equilibrium. 


\bibitem{GWS} A.~Griffin, W.~C.~Wu, and S. Stringari, Phys. Rev. Lett. \textbf{78}, 1838 (1997).


\bibitem{BC} G.~M.~Bruun and C.~W.~Clark, Phys. Rev. Lett. {\bf 83}, 5415 (1999).
 
\bibitem{CG} A.~Csord\'{a}s, R.~Graham, Phys. Rev. A \textbf{64}, 013619 (2001).

\bibitem{griffinAd} T.~Nikuni and A.~Griffin, Phys. Rev. A \textbf{58}, 4044 (1998).


 


\end{thebibliography}
\end{document}